\documentclass{iopart}[12]
\usepackage{mathptmx}
\begin{document}

\renewcommand{\baselinestretch}{0.833}
\newcommand{\beq}{\begin{equation}}
\newcommand{\eeq}{\end{equation}}
\newcommand{\beqar}{\begin{eqnarray}}
\newcommand{\eeqar}{\end{eqnarray}}

\newcommand{\gsim}{\stackrel{>}{\sim}}
\newcommand{\spc}{\mbox{ }}
\newcommand{\dspc}{\mbox{  }}

\title[Modeling the Impulsive Noise Component...]
{Modeling the Impulsive Noise Component and its Effect\protect\\ 
on the Operation of a Simple Coherent Network Algorithm\protect\\
for Unmodeled Gravitational Wave Bursts Detection 
}

\author{Maria Principe and Innocenzo M  Pinto}
\address{WavesGroup, University of Sannio at Benevento, Italy, LSC and INFN}
\eads{\mailto{principe@unisannio.it}, \mailto{pinto@sa.infn.it}}
\date{}
\pacno{04.80.Nn, 05.40.-a, 07.05.Kf, 95.55.Sz}
\begin{abstract}
An analytic model \'{a} la Middleton of the impulsive noise component 
in the data of interferometric gravitational wave detectors is proposed, 
based on an atomic representation of glitches. 
A fully analytic characterization of the coherent network data analysis algorithm 
proposed by Rakhmanov and Klimenko is obtained,
for the simplest relevant case of triggered detection of
unmodeled gravitational wave bursts, using the above noise model. 
The detector's performance is evaluated 
under a suitable central-limit hypothesis,
and the effects of both the noisiness of the {\it pseudo}-templates, 
and the presence of the impulsive noise component are highlighted.
\end{abstract}

\section{Introduction}
\label{intro}
\hspace*{20pt}Gravitational wave (henceforth GW) astronomy 
is expected to open an essentially new observational window on the physical Universe.
Several classes of GWs of cosmic origin are currently
being sought for, including continuous, transient and stochastic ones.
An essential distinction among these different signals 
concerns our ability in modeling the expected waveforms.
GW {\it bursts} (henceforth GWB) 
are a paradigm of transient signals for which 
only a few physically-based models exist
\cite{GWBwfs1}-\cite{GWBwfs3}. 

GW detectors (with specific reference to present-day 
large baseline optical interferometers) 
are invariably affected by transient disturbances 
of various origin \cite{tutorialGlitches}.
Using auxiliary channels to monitor the status 
of the instrument and its environment may help 
identifying and vetoing these disturbances.
Experimental evidence suggests that a residual impulsive component 
will nonetheless be present in the data. 
Distinguishing these spurious noise glitches 
from true GWB of cosmic origin  will  be almost impossible,  
when only data from a single detector are available.
It becomes feasible, in principle, if the outputs of several detectors 
are suitably combined. Using data from several detectors 
it is further possible to reconstruct the GW signal waveform, 
which encodes the relevant source physics,
so as to capitalize on and refine astrophysical models. 

As an historical heritage of {\it acoustic} GW detectors \cite{IGEC}, 
various {\it coincidence} algorithms, based on consistency tests 
among candidate-events gathered by different detectors, 
have been studied and tested \cite{COINC_1}-\cite{COINC_N}.
These algorithms, while conceptually simple and computationally inexpensive, 
turn out to be less efficient, in general, 
compared to {\it coherent} techniques, 
where the output data from several sensors 
are combined  to form a suitable detection statistic
to be used in classical hypotheses tests \cite{cohVScoinc}.
Several coherent techniques have been hitherto proposed
\cite{COH_1}-\cite{RIDGE}, but only a few  
(e.g., WAVEBURST \cite{WaveBurst}, X-PIPELINE \cite{X-PIPELINE} and RIDGE \cite{RIDGE})
have been fully implemented to date in the data analysis pipelines of running experiments. 
Considerable work is still needed 
for completely characterizing alternative coherent algorithms 
in terms of performance and computational cost.

This paper (the first in a suite, where we propose to investigate
problems of increasing complexity) attempts to provide a quantitative answer
to the rather fundamental question of how well a network of several GW detectors
may discriminate {\it true} GWBs from local disturbances ({\it glitches}) 
using {\it coherent} detection statistics. 
Among the essential benefits provided by coherent network operation, 
we mention:  i) the ability of detecting unmodeled signals; 
ii) the  capability of rejecting local disturbances;
the possibility of iii) retrieving the source position on the celestial spere and 
iv) reconstructing the gravitational waveforms.
Here we focus on the first two properties,
assuming for simplicity that the GW direction of arrival (henceforth DOA),
and the time of occurrence (henceforth TOO) of the event 
are known ({\it triggered search}) from observations 
of different nature (e.g., electromagnetic, neutrino, etc.).

To this end, modeling the impulsive noise component (the glitches) 
is a key, and yet open, issue.  
In this paper we adopt, for the first time to the best of our knowledge,
a general representation of impulsive noise proposed by D. Middleton 
in a series of seminal papers \cite{MIDD1}.
Glitches are accordingly modeled 
as time-frequency atoms,  i.e., transients 
whose energy content is almost confined  to a compact region in the time frequency plane,
and characterized in terms of a few (random) parameters;
we adopt the possibly simplest (though observation-driven) model for
such atoms: real-valued sine-gaussian (SG) functions.
The impulsive noise component in each detector,
is modeled as a random train of these atoms,
each occurring independently on the other ones.

In order to keep the formal complexity to a minimum, 
while still capturing the key aspects of the problem, 
we make a number of simplifying assumptions, 
summarized below.

We refer to the coherent cross-correlation method 
introduced by Rakhmanov and Klimenko (henceforth RK) in \cite{RK}
and deduce its statistical properties in analytic form.
This latter may be recognized as a natural extension
of the matched correlator concept to the detection of unmodeled
GWBs with a redundant network of detectors.

We limit to the simplest (though realistic and up-to-date) case 
of a network composed of three interferometers of comparable sensitivity, 
with specific reference to the LIGO-Hanford (henceforth LH), 
LIGO-Livingston (henceforth LL) and Virgo (henceforth V)
detectors, and restrict to the case where the incident wave
is linearly polarized.

This paper is accordingly laid out as follows.
In Section 2 we recall the RK formalism,
and deduce the distribution of its detection statistics 
under the $H_1$ hypothesis (GWB in the data). 
In Section 3 the adopted atomic representation of glitches
is briefly introduced, and the first two moments of the
RK detection statistics under the $H_0$ hypothesis 
(only noise in the data) are derived 
following a simple heuristic reasoning, 
making the simplifying assumption that in each detector
no more than a single glitch may occur in the analysis window. 
This assumption is relaxed in Section 4,
where we propose the anticipated fairly general and rigorous approach 
\'{a} la Middleton to model the impulsive component of the interferometer noise;
the results obtained in Sections 3 and 4 are shown to 
coincide, under the appropriate simplifying assumptions.
In Section 5, based on the above results, 
we evaluate the RK correlator based detector's performance
by numerical experiments. 
Conclusions and hints for future work follow under Section 6.

\section{The RK Coherent Analysis Algorithm}
\label{sec:technsection}

\hspace*{20pt}Whenever the sought waveform  is known a priori, 
optimal detection in additive stationary (band-limited) white gaussian noise 
is achieved by matched-filtering the data with a {\it template} of the sought waveform.
The output of the filter (also known as {\it matched correlator}) 
has to be compared to a properly chosen threshold in order to decide 
about the presence or absence of the signal in the data.
When the signal shape is known, except for a finite number of parameters, 
a set of correlators corresponding to a suitably dense covering of the parameter space 
can be computed, and the largest one exceeding the threshold selected, 
yielding an estimate for the signal parameters.

For unmodeled GWBs the matched filtering technique cannot be adopted. 
It is thus basically impossible to distinguish a GWB 
from a spurious glitch surviving the auxiliary-channel based vetos, 
in the data of a {\it single} interferometer.

One possible way to circumvent this difficulty using data from a (redundant\footnote{
A network of detectors is {\it redundant} 
if, in the absence of noise, the output of each detector can be expressed 
in terms of the outputs of the others.
Since a gravitational signal has only two independent polarization components, 
any network of three or more (differently located and oriented) detectors 
is redundant in the common observational band.})
network of detectors 
has been proposed in \cite{RK}, 
and will be shortly recalled hereinafter.

Let a (plane) gravitational wave 
with linearly polarized (TT gauge) components $h_+(t)$ and $h_\times(t)$  
impinge on a set of interferometric detectors 
located at $\vec{r}=\vec{r}_i$,
$i=1,2,3$, 
from a direction $\Omega_s$.
 
In the absence of noise, the detector outputs can be written\footnote{
In writing eq. (\ref{eq:DetOut}), 
we make the usual (competing) assumptions that 
the gravitational wave signal is i) short enough 
to ignore the reciprocal motion between source
and detector negligible, and  
ii) spectrally narrow  enough 
to make the response of the antenna practically instantaneous.}

\beq
S'_i(t) = F_i^+(\Omega_s)h_+[t-\tau_i(\Omega_s)]+
         F_i^{\times}(\Omega_s)h_{\times}[t-\tau_i(\Omega_s)], 
         \mbox{         }i=1,2,3,
\label{eq:DetOut}
\eeq
where  $\tau_i(\Omega_s)=c^{-1}\widehat{n}\cdot\vec{r}_i$ 
is the propagation delay of the (plane) wavefront 
referred to its arrival at the spatial origin
(usually taken coincident with the Earth center), 
$\widehat{n}$ is the unit wave-vector,  
and  $F_i^+(\Omega_s)$, $F_i^{\times}(\Omega_s)$ 
are the {\it pattern functions}
describing the antenna directional response. 
\\
Equations (\ref{eq:DetOut}) can be rewritten in matrix form
as follows, 

\begin{equation}
\left(
\begin{array}{c}
	S_1(t) \\
	S_2(t) \\
	S_3(t) \\
\end{array}
\right)   =    \left(  \begin{array}{cc}
														F_1^+(\Omega_s) & F_1^{\times}(\Omega_s) \\
														F_2^+(\Omega_s) & F_2^{\times}(\Omega_s) \\
														F_3^+(\Omega_s) & F_3^{\times}(\Omega_s) \\
											 \end{array}
						   \right) 
						   \left(  \begin{array}{c}
						   							h_+(t) \\
						   							h_\times(t) \\
						   				 \end{array}
						   \right),						   
\label{eq:NetOut}
\end{equation}
where

\beq
S_i(t) = S'_i[t+\tau_i(\Omega_s)],\mbox{  }i=1,2,3
\label{eq:delayed}
\eeq
are the properly time-shifted noise-free detector outputs,
and the matrix on the r.h.s.  is the {\it network response matrix}.
The rank of this latter
cannot exceed $2$, hence,

\beq
\mbox{det}
\left(
\begin{array}{ccc}
	S_1 & F_1^+ & F_1^{\times} \\
	S_2 & F_2^+ & F_2^{\times} \\
	S_3 & F_3^+ & F_3^{\times} \\
\end{array}
\right)=0.
\label{eq:determinant}
\eeq
Expanding the determinant in the elements of the first column, 
we obtain the null condition \cite{COH_1}

\beq
A_1(\Omega_s)S_1(t) + A_2(\Omega_s)S_2(t) + A_3(\Omega_s)S_3(t) = 0,
\label{eq:NullCondition}
\eeq
where 
\beq
\left\{
\begin{array}{c}
	A_1 = F_2^+ F_3^{\times} - F_3^+ F_2^{\times} \\
	A_2 = F_3^+ F_1^{\times} - F_1^+ F_3^{\times}. \\
	A_3 = F_1^+ F_2^{\times} - F_2^+ F_1^{\times}
\end{array}
\right.
\label{eq:AAA}
\eeq
In (\ref{eq:AAA}) (and whenever possible, hereafter) 
the dependence on $\Omega_s$ 
of $A_i$ and $F_i^{+,\times}$ is omitted for notational ease.
\\
From eq. (\ref{eq:NullCondition}) one may infer that, 
in the absence of noise, 
the output of each and any detector in the network
is proportional to a linear combination 
of the outputs of the remaining two, 
namely

\beq
\Sigma_i(t) = - A_j S_j(t) - A_\ell S_\ell(t)  = A_i S_i(t),
\mbox{   }i,j,\ell=1,2,3,\mbox{ }j\neq \ell \neq i.
\label{eq:template}
\eeq
and is thus a {\it template}\footnote{
For all practical purposes, one may use as a template for $S_i$ any
quantity differing from $S_i$ by an arbitrary multiplicative factor \cite{Helstrom}.}
for $S_i$. 
The {\it actual} interferometer outputs $V_i(t)$, however, 
differ from the $S_i(t)$ due to the presence of noise, viz. 

\beq
V_i(t) = S_i(t)+n_i(t),\mbox{   }i=1,2,3. 
\label{eq:Vi}
\eeq
Accordingly, by using the $V_i$ in place of $S_i$ in (\ref{eq:template})
we obtain a {\it noisy} template for $S_i$,

\beqar
W_i(t) = 
-  A_j V_j(t) - A_\ell V_\ell(t)  = 
A_i \left[ S_i(t) + \nu_i(t) \right],\nonumber\\
~~~~~~i,j,\ell=1,2,3, \mbox{ }j\neq \ell \neq i,
\label{eq:NoisyTemplate}
\eeqar
where
\beq
\nu_i(t) = - \frac{1}{A_i}\left[ A_j n_j(t) + A_\ell \	n_\ell(t) \right],
\mbox{     }i,j,\ell=1,2,3,\mbox{ }j\neq \ell \neq i.
\label{eq:NoiseInTheTemplate}
\eeq
The $W_i(t)$ can thus be used, following RK \cite{RK}, to compute 
the {\it pseudo} matched-correlators 

\beqar
C_i \equiv \langle V_i,W_i \rangle = 
\int_{\Theta_i} V_i(t)W_i(t) dt \approx\nonumber\\ 
~~~~~~\approx f_s^{-1} \sum_{m=1}^{N_s} V_{im}W_{im} 
\equiv
f_s^{-1} {\bf V}_i \cdot {\bf W}_i^T,\mbox{    }i=1,2,3.
\label{eq:Ci}
\eeqar
In (\ref{eq:Ci}) $\Theta_i$ is  the ($T$ seconds wide) analysis window, 
$f_s$ the sampling frequency,
$N_s=\lfloor f_s T \rfloor$ the number of samples in $\Theta_i$,
$V_{im}$ and $W_{im}$ the time samples of $V(t)$ and $W(t)$, respectively, 
${\bf V}_i = \left\{V_{i1},V_{i2},\dots,V_{iN_s}\right\}$, 
and ${\bf W}_i = \left\{W_{i1},W_{i2},\dots,W_{iN_s}\right\}$. 

In view of the relatively large value of $N_s$ in (\ref{eq:Ci}), typically $\gsim 10^2$, 
in the following we shall make the working assumption
that a suitable form of the (generalized) Central Limit Theorem 
may be invoked \cite{CLT} to argue 
that the $C_i$ are normal distributed under both hypotheses
$H_1$ (signal present) and $H_0$ (no signal),
despite the presence of glitches, which makes the
interferometer noises depart from Gaussianity.
Only the first two moments will be accordingly needed
to characterize them. 

\subsection{RK Correlators Distributions under $H_1$}

\hspace*{20pt}In this Section we assume that the occurrente of a GWB 
{\it and} an instrumental glitch in the {\it same} analysis window 
can be neglected, being extremely unlikely\footnote{ 
This (reasonable) assumption may be relaxed 
using the moments of the total noise 
derived in Sect. 4, in lieu of those 
of the Gaussian component alone.
Results pertaining to this more general
case will be presented elsewhere.}.
Denoting the moments for the $H_1$ case with the subscript 1, 
one readily obtains $(i=1,2,3)$
\beq
\mu_1^{(i)} = E\left[C_i|H_1\right] = A_i(\Omega_s)
\int_{\Theta_i} S_i^2(t) dt
\approx f_s^{-1} 
A_i(\Omega_s) {\bf S}_i \cdot {\bf S}_i^T, 
\label{eq:aveCi}
\eeq
where ${\bf S}_i = \left\{ S_i(t_1),S_i(t_2),\dots,S_i(t_{N_s})\right\}$, and 
\beqar
\left(\sigma_1^{(i)}\right)^2 = Var\left[C_i|H_1\right] = 
A^2_i(\Omega_s)
\left[
\frac{N_i + \tilde{N}_i}{2}
\int_{\Theta_i} S_i^2(t) dt
+\frac{N_i\tilde{N}_i}{4} N_s
			\right]\approx\nonumber\\
~~~~~~\approx			
f_s^{-2}  A_i^2(\Omega_s) \left[ 
\left(
\sigma_i^2 + 
\tilde{\sigma}_i^2
\right)  {\bf S}_i\cdot {\bf S}_i^T + 
\sigma_i^2\tilde{\sigma}_i^2 N_s \right].
\label{eq:varCi}
\eeqar
where $N_i$ and $\tilde{N}_i$ denote 
the one-sided power spectral densities 
of $n_i(t)$ and $\nu_i(t)$ respectively\footnote{
The last term in (\ref{eq:varCi}) is obtained from 
$$
E\left[\int_{[T]}dt\int_{[T]}ds\mbox{  }n_i(t)n_i(s)\nu_i(t)\nu_i(s) \right]
$$
using the band-limited white-noise formula
$$
E[n(t)n(s)]= \frac{N}{2}\int_{-B}^{B} \exp[\imath 2\pi f(t-s)]df.
$$}.
In deriving (\ref{eq:aveCi}) and (\ref{eq:varCi}) we capitalize on 
the statistical independence between $n_i$ and $\nu_i$,  the obvious identities

\beq
E\left[n_i\right] = E\left[\nu_i\right] = E\left[n_i\nu_i\right] = 0,
\eeq
and the relationship $N_i = 2\sigma_i^2/f_s$, 
valid for band-limited gaussian white noise  with standard deviation $\sigma_i$.

The performance of the RK correlator 
is described in terms of its {\it deflection} \cite{Helstrom},
aka signal to noise ratio (SNR) defined as

\beq
d^{(i)} = \frac{\mu_1^{(i)}}{\sigma_1^{(i)}} = 
\frac{ {\bf S}_i \cdot {\bf S}_i^T}
{
\left[ \left( \sigma_i^2 + \tilde{\sigma}_i^2\right) {\bf S}_i \cdot {\bf S}_i^T + 
\sigma_i^2\tilde{\sigma}_i^2 N_s.
\right]^{1/2}},
\label{eq:dsubi}
\eeq
In the following we shall assume for simplicity 
that all detectors in the network have comparable noise PSDs, 
thus letting $N_i = N$ and $\sigma_i = \sigma$, $\forall i$. Accordingly,

\beq
\tilde{\sigma}_i^2 = 
\frac{A^2_{j} + A^2_{\ell} }{A^2_{i}} \sigma^2,
\label{eq:Qi}
\eeq
so that eq. (\ref{eq:dsubi}) becomes

\beq
d^{(i)} = 
\frac{  A_i \left( {\bf S}_i \cdot {\bf S}_i^T \right)^{1/2} }
{\sigma \left[ A_{i}^2 + A_{j}^2 + A_{\ell}^2  + N_s \left( A_{j}^2+A_{\ell}^2 \right)
\displaystyle{   \frac{\sigma^2}{{\bf S}_i \cdot {\bf S}_i^T }   }  
\right]^{1/2}
}.
\label{eq:d_i}
\eeq

The deflection (\ref{eq:d_i}) can be written 
in a more transparent form 
by introducing the quantities

\beq
\delta_{S}^{(i)} = 
\left( \frac{\int_{\Theta_i} S_i^2(t) dt}{N/2} \right)^{1/2} 
\approx f_s^{-1} \frac{\mathbf{S}_i\cdot \mathbf{S}_i^T}{\sigma},
\label{eq:delta_S}
\eeq
representing the signal to noise ratio 
of a {\it perfect} matched filter applied to the
{\it actual} data at the output of detector-$i$, and

\beq
\delta_h = \left(
\frac{h_{rss}^2}{N/2} 
\right)^{1/2},
\label{eq:delta_h}
\eeq
where
\beq
h_{rss}^2 = \int_{\Theta_i} \left[h_+(t)^2 + h_{\times}(t)^2 \right] dt
\eeq
is a frequently used measure of the GWB strength.
For the simplest case of linearly polarized GWBs, 
\beq
\delta^{(i)}_S = \left|F_i\right| \delta_h, 
\label{eq:deltaS_vs_delta_h}
\eeq
where $F_i=F^{+,\times}_i$, depending on the wave polarization, 
and $\delta_h$  represents the {\it intrinsic} signal to noise ratio 
of a perfect matched filter applied to the {\it bare} gravitational waveform 
embedded in the detector noise; such a deflection would be attained 
if the antenna were isotropic $(F^{+,\times}_i = 1)$, 
and the template noise-free. 

The deflection $d^{(i)}$ in (\ref{eq:dsubi}) 
can thus be conveniently written:
\beq
d^{(i)} = \delta_S^{(i)}\Xi_i(\Omega_S,N_s,F_i,\delta_h),
\eeq
where 
\beq
\Xi_i(\Omega_S, N_s,F_i,\delta_h)=
\displaystyle{
\frac{A_i}{
\left[ A_{i}^2 + A_{j}^2 + A_{\ell}^2  + N_s \left( A_{j}^2+A_{\ell}^2 \right)
( \left|F_i\right| \delta_h )^{-2}
\right]^{1/2}}}.
\label{eq:Xi}
\eeq
measures the SNR degradation of the RK correlator $C_i$ w.r.t. 
the {\it perfect} matched filter acting on the output data of detector-i, 
due to the noisiness of the template.
It is seen from (\ref{eq:Xi}) that $\Xi_i(\cdot)$ depends on the DOA, 
the number of samples $N_s$ in the analysis window,
the polarization-dependent pattern function $F_i$, 
and the intrinsic deflection $\delta_h$.

Figures 1a and 1b display the sky maps of the function 
$\Xi_i(\cdot)$ in (\ref{eq:Xi})
for the LH, LL and V detectors 
for the two linear polarizations,
for $N_s=100$ and two extremal values of $\delta_h$, 
namely $\delta_h=10$ and $\delta_h=100$, respectively. 
The source position in Figs. 1a and 1b is parameterized
in terms of the polar and  azimuthal angles 
$\vartheta_s,\varphi_s$ in an Earth-centered 
coordinate system whose polar axis points to the North-Pole,
and $\varphi_s=0$ identifies the Prime Meridian.

\section{Glitches}
\label{sec:burstmodel}

\hspace*{20pt}Available experimental evidence  \cite{Soma},\cite{Saulson}
suggests that instrumental noise glitches 
can be efficiently modeled as {\it atoms} \cite{Gabor} in the time-frequency plane\footnote{
GWBs can also be modeled as atoms. For the detection technique
adopted here, however, only the GWB energy is relevant, 
{\it not} its shape.}
\cite{SuttonCheese}.
Atoms are waveforms with almost-compact time-frequency support. 
They can be characterized in terms of their energy content, 
and their first and second order moments, i.e., 
occurrence time $t_0$, center frequency $f_0$, 
effective duration $\sigma_t$ and bandwidth $\sigma_f$.

The choice of an atom family 
(technically called a {\it dictionary}) appropriate to modeling
glitches in GW interferometers must be 
compliant to and derived from experimental evidence.
In this connection, the work in \cite{Soma},\cite{Saulson},
aimed at classifying glitches and identifying glitch 
families (clusters in parameter space) is particularly relevant. 
It should be noted that atoms in general form {\it overcomplete} systems, 
and this fact must be taken properly into account in deducing
the distributions of the atom parameters from
observed glitch populations \cite{MallatZhang}.

We adopt here the possibly simplest atom,  
the real valued sine-Gaussian (SG) functions defined by
\beq
\psi(t-t_0;g_0,f_0,\sigma_t)=g_0 \sin\left[2\pi f_0 (t-t_0)\right] e^{-(t-t_0)^2/\sigma_t^2},
\label{eq:SG}
\eeq
whose waveform and time-frequency representation are shown in Figure 2.
The SG atom is entirely characterized by its 
{\it shape} parameters $g_0$, $f_0$ and $\sigma_t$, 
and effective occurrence (firing) time $t_0$.

The choice of the SG dictionary is suggested 
by the fact that a wide variety of observed glitches 
in the data channel are well modeled as SG atoms \cite{Saulson},
and is further motivated by its structural simplicity, 
minimum time-frequency spread, 
($\sigma_t \sigma_f = (4\pi)^{-1}$),
and positive-definiteness of its Wigner-Ville transform.
These properties should likely permit to represent 
the instrumental transients in a close-to-optimal 
(i.e., minimally redundant) way (see, e.g., \cite{HelstromGaussExp}, \cite{BastiaanGaussExp}).

\subsection{RK Correlators Distributions under $H_0$ - Heuristic Approach}  
\label{sec:H0_heuristics}

\hspace*{20pt}In this Section we obtain a heuristic characterization of
the RK correlator distribution under $H_0$ by considering the glitches 
as (spurious) signals with random parameters. 
Specifically, we derive the "average" among the marginal distributions
of the RK correlator corresponding to all possible glitch realizations 
in the network. To keep the analysis as simple as possible, we assume 
that no more than a single glitch may occur in the analysis window
in each interferometer.
This restriction will be removed in the next Section, 
where a fairly general and rigorous model 
for the impulsive noise component will be proposed.

It is expedient to write the moments under $H_0$
(denoted with a subscript $0$) as follows:

\beq
\mu_0^{(i)} = (1-\Pi)^3\mu_{0,0}^{(i)} + \Pi(1-\Pi)^2\mu_{0,1}^{(i)} + \Pi^2(1-\Pi)\mu_{0,2}^{(i)} + \Pi^3\mu_{0,3}^{(i)},
\label{eq:mu0}
\eeq
\beqar
\left(\sigma_0^{(i)}\right)^2 = (1-\Pi)^3\left(\sigma_{0,0}^{(i)}\right)^2 + \Pi(1-\Pi)^2\left(\sigma_{0,1}^{(i)}\right)^2 + \nonumber\\
~~~~~~~~+\Pi^2(1-\Pi)\left(\sigma_{0,2}^{(i)}\right)^2 + \Pi^3\left(\sigma_{0,3}^{(i)}\right)^2,
\label{eq:sig0}
\eeqar
where $\mu_{0,k}^{(i)}$, $\sigma_{0,k}^{(i)}$ refer to the cases where $k$ detectors
($k$ = 0,1,2,3) in the network exihibit a glitch within the analysis window, and
the corresponding factors in front of them are the related occurrence probabilities,
$\Pi$ being the (known) probability of observing a single glitch in the analysis window.

The quantities $\mu^{(i)}_{0,k}$ and $\sigma^{(i)}_{0,k}$ 
in eq.s (\ref{eq:mu0}) and (\ref{eq:sig0}) can be computed for any glitch 
{\it instance} in the network, i.e. for any allowed set of (possibly null) 
SG atoms in the interferometers' outputs.
For each instance, these quantities identify the corresponding {\it marginal} moments
of the detection statistics $C_i$ under $H_0$. 
We are obviously interested in computing the same moments averaged over 
{\it all} possible glitch realizations 
in the network detectors, using the known prior distributions
for the glitch parameters.
After some tedious algebra, we accordingly get (under
the usual discrete-time representation):

\beq
\begin{array}{l}
\mu_{0,0}^{(i)} = 0,\\
\left(\sigma_{0,0}^{(i)}\right)^2 = 
 f_s^{-2}(A^2_j+A^2_\ell)\sigma^4N_s,
\end{array}
 \label{eq:sig00}
\eeq

\beq
\begin{array}{l}
\mu_{0,1}^{(i)} = 0,\\
\left(\sigma_{0,1}^{(i)}\right)^2 = 3\left(\sigma_{0,0}^{(i)}\right)^2 + 
										2f_s^{-2}\sigma^2(A^2_j+A^2_\ell)E({\mathbf \psi}\cdot {\mathbf \psi}^T),
\end{array}
\eeq

\beq
\begin{array}{l}
\mu_{0,2}^{(i)} = -f_s^{-1}(A_{j}+A_{\ell})E({\mathbf \psi}\cdot{\mathbf \psi'}^T),\\
\left(\sigma_{0,2}^{(i)}\right)^2 = 3\left(\sigma_{0,0}^{(i)}\right)^2 + 
										f_s^{-2}\sigma^2\left[
											4(A^2_j+A^2_\ell)E({\mathbf \psi}\cdot{\mathbf \psi}^T) +
											2A_{j}A_{\ell}E({\mathbf \psi}\cdot{\mathbf \psi'}^T) \right],
\end{array}
\eeq

\beq
\begin{array}{l}
\mu_{0,3}^{(i)} = -f_s^{-1}(A_{j}+A_{\ell})E({\mathbf \psi}\cdot{\mathbf \psi'}^T),\nonumber\\
\left(\sigma_{0,3}^{(i)}\right)^2 = \left(\sigma_{0,0}^{(i)}\right)^2 + 
										f_s^{-2}\sigma^2\left[
											2(A^2_j+A^2_\ell)E({\mathbf \psi}\cdot{\mathbf \psi}^T) +
											2A_{j}A_{\ell}E({\mathbf \psi}\cdot{\mathbf \psi'}^T) \right],
\end{array}
\label{eq:sig03}
\eeq
where ${\mathbf \psi}=\left\{\psi(t_1),\psi(t_2),\dots,\psi(t_{N_s}) \right\}$, and
$E({\mathbf \psi}\cdot{\mathbf \psi}^T)$ and $E({\mathbf \psi}\cdot{\mathbf \psi'}^T)$ 
(multiplied by $f_s^{-1}$) are the expected glitch energy 
and the expected correlation between glitches 
occurring in different detectors, respectively, 
both expectations being taken over all possible glitch instances.

Note that eqs. (\ref{eq:sig00}) are nothing but the moments of $C_i$ 
under $H_0$ in the absence of glitches, i.e., due to the Gaussian noise
floor only. 
It is therefore apparent that glitches have a twofold effect,
making the expected value of $C_i$ non-zero, 
and increasing its variance.

Substituting eqs. (\ref{eq:sig00})-(\ref{eq:sig03}) into eqs. (\ref{eq:mu0}) and (\ref{eq:sig0}),
the (marginalized) first two moments of the
detection statistic under $H_0$ can be written ($i = 1,2,3$)

\beq
\mu_0^{(i)} = -\Pi^2f_s^{-1}\left(A_{j}+A_{\ell}\right)E({\mathbf \psi}\cdot{\mathbf \psi'}^T),
\label{eq:muHGheu}
\eeq
and
\beq
\left(\sigma_{0}^{(i)}\right)^2 = \left(\sigma_{0,0}^{(i)}\right)^2
					\left\{ 1+ \frac{2}{N_s} \left[
					\Pi\frac{E({\mathbf \psi}\cdot{\mathbf \psi}^T)}{\sigma^2} + 
					\Pi^2{\cal H}(\Omega_s)\frac{E({\mathbf \psi}\cdot{\mathbf \psi'}^T)}{\sigma^2}\right] \right\},
\label{eq:sig2HGheu}
\eeq
where

\beq
{\cal H}(\Omega_s) = \frac{2A_{j} A_{\ell}}{(A^2_j+A^2_\ell)}.
\eeq

\section{The Impulsive Component - Toward a Rigorous Approach}
\label{sec:Middleton_model}

\hspace*{20pt}In this Section we relax the assumption 
made in the previous Section
that no more than a single glitch 
may occur in the analysis window in each interferometer, 
and adopt a general, fully rigorous approach 
to model the glitchy component,
along the lines laid out by D. Middleton in a series of seminal papers \cite{MIDD1}.

The impulsive noise component in each interferometer 
is  accordingly modeled as a {\it random process} 
consisting of a linear superposition of atoms, viz.\footnote{ 
A straightforward generalization is obtained by adding 
several terms like (\ref{eq:H0Process}) using
{\it different} atom families.
The characteristic function of the resulting process 
will be the product of the characteristic functions of its terms.}

\beq
g_i(t) = \sum_{k=1}^{K_i[T]} \psi\left(t-t_{0,i}^{(k)};\vec{a}_i^{(k)}\right), \mbox{   }t \in \Theta_i,
\mbox{   }i=1,2,3.
\label{eq:H0Process}
\eeq
Here $\psi(\cdot)$ is the chosen representation atom,
$t_{0,i}^{(k)}$ are a set of random glitch firing times, 
$\vec{a}_i^{(k)}$ is a set of random (independent) shape-parameters 
(e.g., amplitude, center frequency, duration, bandwidth),
and $K_i[T]$ is also a random variable, denoting the number of glitches 
occurring in the analysis window $\Theta_i$, whose time-width is denoted as $T$.
The firing-times and shape-parameters are determined independently 
at each glitch occurrence (i.e., for each $k$).

The key modeling assumption is that
the glitching component may be taken as stationary (homogeneous) 
on time scales sufficiently long compared to the analysis window. 
On such time scales {\it typical} glitches will show up 
with {\it constant} probabilities,
and occur at a {\it constant} rate,
which can be stimated from actual data.

The number of events  $K_i[T]$ will be accordingly ruled 
\cite{HurwitzKac} by a Poisson distribution, i.e.

\beq
\mbox{prob}\left\{K_i[T]=K_i \right\} = \frac{\overline{N}_i^{K_i} e^{-\overline{N}_i}}{K_i!},
\eeq
where $\overline{N}_i$ (aka, $\lambda_i T$, $\lambda_i$ being the glitch firing-rate) 
is the average number of glitches occurring in interferometer-$i$ 
in the ($T$-seconds wide) analysis window.
We shall assume the above probability, as well as the distributions
of the firing-times and shape-parameters in (\ref{eq:H0Process}), to be the same 
for all instruments, and henceforth drop the index $i$.

The characteristic functions of the process (\ref{eq:H0Process})
can be computed exactly
up to any order \cite{MIDD1}. 
The first order one can be written

\beq
F_{g}(\xi,t) = \sum_{K=0}^{\infty}\mbox{prob}\left\{K[T]=K\right\} F_{g}(\xi,t|K),
\label{eq:Fxit}
\eeq
where $F_{g}\left(\xi,t|K\right)$ is the conditional characteristic function, given $K$
glitches in the analysis window $\Theta$, viz.:
\beq
F_{g}(\xi,t|K) = 
E\left\{ 
\exp
\left[ 
\imath \xi 
\displaystyle{											
\sum_{m=1}^{K} 
\psi\left(t-t_0^{(m)};\vec{a}^{(m)}\right)} 
\right] 
\right\}.
\label{eq:FxitK}
\eeq
The expectation  in (\ref{eq:FxitK}) is taken with respect to 
both the firing times, $t_0^{(m)}$, and the shape parameters, $\vec{a}^{(m)}$.
The pertinent distributions being assumed as time-invariant in $\Theta$,
and independent for each glitch occurrence, eq. (\ref{eq:FxitK}) 
and (\ref{eq:Fxit}) become, respectively

\beq
F_{g}(\xi,t|K) = 
E\left[ 
e^{j\xi \psi(t-t_0;\vec{a})} 
\right]^{K},
\eeq
and

\beq
F_{g}\left( \xi,t \right) = 
\exp
\left[
\overline{N}
\left(
E\left[ 
        e^{j\xi \psi(t-t_0;\vec{a})} 
 \right]-1 
\right) 
\right].
\label{eq:FxitS}
\eeq

From the characteristic function $F_{g}\left( \xi,t \right)$
it is straightforward to compute the moments of the process $g(t)$,
representing the impulsive (glitch) noise component in each
interferometer:

\beq
\mu_{g}^{(Q)} = 
(-\imath)^{Q}
\left.
\frac{
\partial^{Q}F_{g}
\left( \xi,t \right)
}{
\partial \xi^{Q}
} 
\right|_{\xi=0},
\eeq
yielding
\beq
\mu_{g}^{(1)} = E[g(t)] = \overline{N}E[\psi(t-t_0;\vec{a})],
\label{eq:muG}
\eeq
and
\beq
\mu_{g}^{(2)} = E[g(t)^2] = \overline{N}^2 E^2[\psi(t-t_0;\vec{a})] + \overline{N}E[\psi^2(t-t_0;\vec{a})],
\label{eq:sigG}
\eeq
where, the expectations are taken with respect to both $t_0$ and $\vec{a}$.
The related distributions being assumed as time-invariant in $\Theta$,
the moments (\ref{eq:muG}), (\ref{eq:sigG}) are also time-independent.

\subsection{RK Correlators Distributions under $H_0$ - Rigorous Approach}
\label{sec:H0_rigorous}

\hspace*{20pt}Using the model exploited in Section \ref{sec:Middleton_model} 
for the impulsive component of the instrument noise, 
it is possible to compute the first two moments of the distribution of $C_i$
under $H_0$ in a rigorous way.
Formally, these are obtained by making the substitution

\beq
n(t) \longrightarrow n(t)+g(t),
\eeq
for the noise in each detector
in computing $E[C_i|H_0]$ and $Var[C_i|H_0]$, thus obtaining
(to second order in the noise moments of $g$)
\beq
\begin{array}{l}
\mu_0^{(i)} = E[C_i|H_0] = -f_s^{-1} N_s (A_j+A_{\ell}) \left(\mu_g^{(1)}\right)^2,
\end{array}
\label{eq:muCiHG0}
\eeq
and
\beq
\begin{array}{l}
(\sigma_0^{(i)})^2 = Var[C_i|H_0] = (A^2_j+A^2_\ell) \sigma^4f_s^{-2}N_s+\\
~~~~~~+ \sigma^2f_s^{-2}N_s 
        \left\{ 2(A^2_j+A^2_\ell) \mu_g^{(2)} + 
				2A_j A_{\ell} \left(\mu_g^{(1)}\right)^2 \right\}=\\
~~~~~~=\left(\sigma_{(0,0)}^{(i)}\right)^2
\left\{1+2\left[\displaystyle{
\frac{\mu_g^{(2)}}{\sigma^2}+
{\cal H}(\Omega_s)\frac{\left(\mu_g^{(1)}\right)^2}{\sigma^2}
}\right]\right\},
\end{array}
\label{eq:sig2CiHG0}
\eeq
and then using eq.s  (\ref{eq:muG}) and (\ref{eq:sigG}) for the first two moments
of the impulsive components $g_i(t)$ in (\ref{eq:muCiHG0}), (\ref{eq:sig2CiHG0}) to get: 

\beq
\begin{array}{l}
\mu_0^{(i)} = E[C_i|H_0] = -f_s^{-1} N_s (A_j+A_{\ell}) \bar{N}^2 E^2[\psi(t-t_0;\vec{a})],
\end{array}
\label{eq:muCiHG}
\eeq
and
\beq
\begin{array}{l}
(\sigma_0^{(i)})^2 = Var[C_i|H_0] = \left(\sigma_{(0,0)}^{(i)}\right)^2
\left\{1+2\left[
\displaystyle{
\bar{N}\frac{E[\psi^2(t-t_0;\vec{a})]}{\sigma^2}+
}\right.\right.\\
~~~~~~~~~~~~~~~~~~~~~~~~~~~~~~+\left.\left.\displaystyle{
\left[1+{\cal H}(\Omega_s)\right]
\bar{N}^2\frac{E^2[\psi(t-t_0;\vec{a})]}{\sigma^2}
}
\right]
\right\}.
\end{array}
\label{eq:sig2CiHG}
\eeq

It is now interesting to compare eq.s (\ref{eq:muCiHG}) and (\ref{eq:sig2CiHG})
to eq.s (\ref{eq:muHGheu}) and (\ref{eq:sig2HGheu}), obtained from the heuristic reasoning in the previous Section.
In order to do so, the sum in (\ref{eq:Fxit}) should include only the $K=0,1$ terms,
to match the assumption made there 
that no more than a single glitch may occur in the analysis window.
This gives the following approximate expressions for the first two moments 
of the impulsive noise component,

\beq
E[g(t)|K=0,1] = \overline{N}e^{-\overline{N}}E[\psi(t-t_0;\vec{a})],
\eeq
\beq
E[g^2(t)|K=0,1] = \overline{N}e^{-\overline{N}}E[\psi^2(t-t_0;\vec{a})],
\eeq
yielding, upon substitution in (\ref{eq:muCiHG0}) and (\ref{eq:sig2CiHG0}),

\beq
E[C_i|H_0,K=0,1] = -f_s^{-1} N_s (A_j+A_{\ell})\overline{N}^2e^{-2\overline{N}}E^2[\psi(t-t_0;\vec{a})],
\label{eq:muCiK1}
\eeq
and

\beqar
Var[C_i|H_0,K=0,1] = \left(\sigma_{(0,0)}^{(i)}\right)^2
\left\{1+2\left[\displaystyle{
\frac{\bar{N} e^{-\bar{N}} E[ \psi^2(t-t_0;\vec{a}) ]}{\sigma^2}
}\right.\right.+\nonumber\\
~~~~~~~~~~~~~~~~~~~~~~~~~~~~~~+\left.\left.\displaystyle{
{\cal H}(\Omega_s)\frac{\bar{N}^2e^{-2\bar{N}} E^2[\psi(t-t_0;\vec{a})]}{\sigma^2}
}\right]\right\}.
\label{eq:sig2CiK1}
\eeqar
Equations (\ref{eq:muCiK1}), (\ref{eq:sig2CiK1}) reproduce
eqs. (\ref{eq:muHGheu}), (\ref{eq:sig2HGheu}) iff 

\beq
\left.
\begin{array}{l}
\Pi N_s^{-1}E({\mathbf \psi}\cdot{\mathbf \psi}^T) = \overline{N} e^{-\overline{N}} E[\psi^2(t-t_0;\vec{a})],\\
\Pi^2 N_s^{-1}E({\mathbf \psi}\cdot{\mathbf \psi'}^T) = \overline{N}^2 e^{-2\overline{N}} E^2[\psi(t-t_0;\vec{a})].
\end{array}
\right.
\label{eq:equality2}
\eeq
Both equalities are trivially proven, noting that for Poissonian distributions
$\Pi=\bar{N}e^{-\bar{N}}$.

In conclusion, the rigorous approach sketched above agrees, in the appropriate limit,
with the result obtained in the previous Section from a simple heuristic argument.
The rigorous approach, on the other hand, allows {\it any} number of glitches 
in the analysis window in each interferometer, in a natural way.

For the special case of SG atoms, assuming a uniform distribution 
of the glitch firing time over the analysis window, we have

\beq
\mu_g^{(1)} = 0,\mbox{     }
\mu_g^{(2)} =  \frac{ \bar{N} \sqrt{\pi} }{2 \sqrt2 T}E\left[g_0^2 \sigma_t \left( 1-e^{-2\pi^2f_0^2\sigma_t^2} \right) \right].
\label{eq:gMomSG}
\eeq

The analytic results for the moments 
in eqs. (\ref{eq:muCiHG0}), (\ref{eq:sig2CiHG0}), (\ref{eq:gMomSG})
were checked successfully against Monte Carlo simulations.

\section{RK Correlator Based Detector Performance}

\hspace*{20pt}Under the made assumption of Gaussianity of the distributions of the $C_i$
under both $H_1$ and $H_0$, it is straightforward to obtain the 
Receiver Operating Characteristics (ROCs) \cite{Helstrom}, which
completely characterize the RK-correlator based detector.

In the appropriate {\it surveillance} context, the detection thresholds
are determined, according to the Neyman-Pearson criterion,
from the prescribed false alarm probability $\alpha$ as 

\beq
\gamma^{(i)} = \sigma_0^{(i)}\mbox{erfc}^{-1}(\alpha) + \mu_0^{(i)},
\mbox{ }i=1,2,3,
\label{eq:thresholds}
\eeq
Note that, in view of eqs. (\ref{eq:muCiHG}) and (\ref{eq:sig2CiHG}),
the thresholds depend on the variance of the Gaussian noise floor,
the DOA, the number of samples in the analysis window,
and, in view of eqs. (\ref{eq:muG}), (\ref{eq:sigG}),
the average glitch energy and firing rate.

The corresponding false dismissal probabilities are

\beq
\beta^{(i)} = 1-P_D^{(i)} = 1-\mbox{erfc}\left(\frac{\gamma^{(i)}-\mu_1^{(i)}}{\sigma_1^{(i)}} \right).
\mbox{ }i=1,2,3,
\label{eq:betas}
\eeq
Combining equations (\ref{eq:thresholds}) and (\ref{eq:betas}), 
we get the explicit expression of the ROCs

\beq
P_D^{(i)} = \mbox{erfc}\left[\frac{\sigma_0^{(i)}}{\sigma_1^{(i)}} \mbox{erfc}^{-1}(\alpha) + \frac{\mu_0^{(i)}}{\sigma_1^{(i)}} -\frac{\mu_1^{(i)}}{\sigma_1^{(i)}} \right],
\mbox{ }i=1,2,3
\eeq
where $P_D^{(i)}$ is the detection probability.

In the numerical experiments illustrated below 
we consider the three-detectors network 
including  LH, LL and V, using these subscripts accordingly.

Equations (\ref{eq:thresholds}) and (\ref{eq:betas}) have been used
to obtain the ROCs ($\beta$ vs $\alpha$ curves\footnote{
Strictly speaking, these are not ROCs, according to the usual definition, 
although ROCs can be trivially derived from them.})
shown in Figures 3 to 8.
All figures refer to the $+$ polarized case.  
Similar results are obtained for the $\times$ polarized case,
and are not reported for brevity.

It is interesting first to illustrate  
how the mere noisiness of the pseudo-templates (\ref{eq:NoisyTemplate}) 
spoils the performance  of the RK detector 
compared to the perfect matched filter. 
To do so, we shall momentarily ignore the impulsive noise component,
by letting $\bar{N}=0$.

Figure 3 shows the ROCs of the RK correlators (best DOA assumed) 
for different values of the {\it intrinsic} SNR $\delta_h$ of the GWB.
The ROC for the perfect matched filter corresponding to  $SNR=7$, 
which is conventionally considered 
as the lowest operational value for this latter,
is also shown.
The value $\delta_h=7$ corresponds, in the pertinent best DOAs, 
to $\delta_S=$6.22 (LH), 6.48 (LL), 6.68 (V).

It is seen that the RK detector's performance  is better when using
$C_{LH}$ or $C_{LL}$ as a detection statistic. In this case, at a given false-alarm rate,
one needs roughly to double the SNR to obtain the same false-dismissal level 
as the perfect matched filter.
Using $C_V$, instead, the SNR must be larger by  a factor $\sim 4$. 
This is due to the different directional response
of Virgo compared to the two LIGOs, 
due to its different orientation.
This is illustrated in Table I, where we collected the values of the quantities

\beq
\bar{\rho}^{(i)}=\frac{\delta_S^{(i)}}{\delta_h}=|F_+^{(i)}|, \mbox{  }
\bar{\rho}_T^{(i)}=\frac{\delta_T^{(i)}}{\delta_h} = \frac{\left|F_i A_i\right|}{\left(A^2_j+A^2_{\ell}\right)^{1/2}},
\label{eq:SNR_in_data_and_pseudotemplate}
\eeq
representing the signal to noise ratios, 
normalized to the {\it intrinsic} SNR of the incoming GWB,
of the data ($\bar{\rho}$) and the noisy template ($\bar{\rho}_T$), respectively.
Table-I shows that in the DOA ranges where Virgo 
exhibits the largest response (largest normalized SNR $\bar{\rho}$), 
the two LIGOs respond poorly, 
and the pseudo-template obtained from them 
has a low normalized SNR $\bar{\rho}_T$. 
Conversely, in the DOA ranges where either of the LIGOs
has the largest response (largest $\bar{\rho}$),
the pseudo-template constructed from the other LIGO and Virgo
has still a decent normalized SNR $\bar{\rho}_T$.   

Figure 4 shows the ROCs for $C_{LH}$ for $\delta_h=10$ and 
three typical durations of the analysis window ($20$, $40$ and $100$ $ms$). 
The ROCs for $C_{LL}$ and $C_V$ are similar, and are not shown for brevity.

Figure 5 is the same as Figure 3, except that here the performances 
are averaged over the whole celestial sphere. 
In this case, the RK correlator based detector performs worse 
than the perfect matched filter roughly by a factor of 3
in terms of SNR.

The effect of instrumental glitches is illustrated in Figures 7 and 8.

In order to draw these figures, we estimated the parameter distributions 
of the SG-atoms to be used  in (\ref{eq:H0Process}) 
from (unclustered) triggers collected in 1 week of S5 data,
kindly provided  by  S. Chatterji \cite{Shourovwebpage}.
The distributions obtained for $f_0$ and $\sigma_t$ are sketched in Figure 6. 
The SG-atom amplitude distribution was assumed as uniform
in an interval set by the maximum SNR in each detector,
beyond which the data are vetoed-out, denoted
as $SNR_{veto}$. 

In Figure 7, the ROCs for $C_{LH}$, $C_{LL}$, and $C_V$ 
are shown for different values 
of the glitch firing rate $\lambda$, 
and compared to the no-glitch case.
Obviously, as $\lambda$ increases, the best achievable
false-dismissal vs. false-alarm probability trade-off deteriorates.
Here the glitch amplitude si assumed as being uniformly
distributed up to level corresponding to $SNR_{veto}=100$.

Finally, in Figure 8 the way the chosen  $SNR_{veto}$ value 
affects the performance is illustrated.

\section{Conclusions and Directions for Future Work}

\hspace*{20pt}We modeled the impulsive noise component following Middleton, 
using an atomic representation for the glitch population. 
The proposed model allows to describe analytically the detector's
performance in the presence of glitches.

Based on the above, we also presented a simple, 
fully analytic characterization 
of the RK coherent network data analysis algorithm
for detecting unmodeled GWBs with known DOA and TOO.
Under a reasonable central-limit hypothesis
for the RK detection statistics distributions,
we derived and discussed the detector's performance, 
in terms of its operating characteristics. 

Our main results can be summarized as follows. 
The presence of noise in the pseudo-templates
spoils the deflection, compared to a perfect matched filter.
The related degradation factor depends on the direction of arrival, 
the energy of the signal, and the length of the analysis window.
The detection threshold, on the other hand, depends on
the variance of the Gaussian noise floor,
the DOA, the number of samples in the analysis window,
the average glitch firing rate, and the maximum allowed
(veto dependent) glitch energy.
Constant False Alarm Rate (CFAR) operation is possible, 
and the RK detector turns out to be reasonably robust 
against instrumental/environmental glitches.

More or less straightforward developments of this work include
i) using a better detection statistic, e.g., a linear combination 
of the RK correlators with (DOA-dependent) coefficients 
chosen so as to maximize the deflection,
and ii) allowing for a  time-varying glitch firing-rate 
(Cox processes \cite{CoxProc}).

As possible directions for future work we mention i) identifying a better atom dictionary, 
and characterising more accurately the prior distributions 
of the relevant parameters using a systematic matching-pursuit based analysis \cite{MallatZhang} 
of the available glitch databases, and 
ii) exploiting in full Middleton's model  
to derive more efficient implementations of the detector. 
In this connection, we note that the straightforward extension 
of the matched correlator to unmodeled waveforms provided 
by the RK algorithm is likely to be {\it not} optimal 
in view of the {\it non}-Gaussian nature of the instruments noise,
whereby some suitable pre-conditioning of the data
will be most likely required  \cite{MIDDBClass}.

Finally, we mention the possibility of integrating Middleton's model 
in a full-fledged interferometer noise simulator including glitches. 
Work along these directions is in progress.

\section*{Acknowledgements}

\hspace*{20pt}We thank  V. Pierro for having introduced to us Middleton's model.
We are also indebted to L. Cadonati, S. Chatterji, E. Katsavounidis, S. Klimenko, 
P. Shawan, P. Saulson, P. Sutton and the LIGO Burst Working Group 
for encouragement and help. We also thank the anonymous Referees for
several suggestions made in their reviews.
  
\newpage
\begin{center}
\Large
Captions to the Figures
\end{center}
\normalsize
$$\mbox{  }$$

Figure 1a - Sky maps of the factor $\Xi_i$ (eq. (\ref{eq:Xi})) 
for the RK correlators $C_i$ corresponding to LH (left), LL (mid) and V (right),
for $+$ (top), and $\times$ (bottom) linear polarizations. 
$T$ = 40 ms; $f_s = 4096$ Hz; $\delta_h = 10$. All panels, 
X axis : $\phi_s$ [rad]; Y axis : $\theta_s$ [rad].\\
$$\mbox{  }$$

Figure 1b - Sky maps of the factor $\Xi_i$ (eq. (\ref{eq:Xi}))
for the RK correlators $C_i$ corresponding to LH (left), LL (mid) and V (right),
for $+$ (top), and $\times$ (bottom) linear polarizations.  
$T$ = 40 ms; $f_s = 4096$ Hz; $\delta_h = 100$. All panels, 
X axis : $\phi$ [rad]; Y axis : $\theta$ [rad].\\
$$\mbox{  }$$

Figure 2 - SG atom with $g_0 = 1$, $t_0 = 0.5$ s, $f_0 = 100$ Hz, $\sigma_t = 0.02$ s. 
Top: time domain waveform; bottom: time-frequency (Wigner-Ville) representation.\\
$$\mbox{  }$$

Figure 3 - Performance in terms of ROCs of the RK pseudo-correlator.  
Optimal DOAs. $T=100$ ms, $f_s=4096$ Hz, several $\delta_h$ values. 
Left: $C_{LH}$; mid: $C_{LL}$; right: $C_V$.\\
$$\mbox{  }$$

Figure 4 - Performance in terms of ROCs of the RK pseudo-correlator.
Optimal DOAs. $f_s=4096$ Hz; $\delta_h = 10$; several $T$. $C_{LH}$ only. \\
$$\mbox{  }$$

Figure 5 - All-sky averaged performance in terms of ROC curves of the RK pseudo-correlator.
$T=100$ ms; $f_s=4096$ Hz. Left: $C_{LH}$; mid: $C_{LL}$; right: $C_V$.\\
$$\mbox{  }$$

Figure 6 - Histograms of SG-atom parameter distributions from 1 week of S5 data \cite{Shourovwebpage}). 
Left: center frequency ($f_0$); right: effective duration ($\sigma_t$).\\
$$\mbox{  }$$

Figure 7 - Performance in terms of ROCs of the RK pseudo-correlator for different glitch rates. 
Optimal DOAs. $T=100$ ms; $f_s=4096$ Hz; $\delta_h=15$; $SNR_{veto}=100$. Left: $C_{LH}$; mid: $C_{LL}$; right: $C_V$.\\
$$\mbox{  }$$

Figure 8 - Performance in terms of ROCs of the RK pseudo-correlator for different $SNR_{veto}$ levels.
Optimally oriented source. $T=100 ms$; $f_s=4096$ Hz; $\delta_h = 15$; $\bar{N}=0.1$. $C_{LH}$ only.\\
$$\mbox{  }$$

Table I - The quantities in (\ref{eq:SNR_in_data_and_pseudotemplate}) evaluated at optimal DOAs.

\section*{References}

\end{document}